\documentclass[aps,pra,superscriptaddress,showkeys,10pt]{revtex4-2}

\usepackage{amsfonts}
\usepackage{graphicx}
\usepackage{amssymb}
\usepackage{amsmath}

\begin{document}

\title{Effect of the odd-channel of a pair of different bosons emerging in multi-species systems and the diffusion of a good quantum number}

\author{Yanzhang He}
\email{Corresponding author: heyanzh@mail.sysu.edu.cn}
\affiliation{School of Physics, Sun Yat-Sen University, Guangzhou, 510275, P. R. China}

\author{Yimin Liu}
\affiliation{School of Intelligent Engineering, Shaoguan University, Shaoguan, 510205, P. R. China}

\author{Chengguang Bao}
\affiliation{School of Physics, Sun Yat-Sen University, Guangzhou, 510275, P. R. China}

\date{\today}

\begin{abstract}
It is well known that, when the Hamiltonian of a quantum mechanic system contains commutable terms, the eigen-states are specified by a set of good quantum numbers.
However, when the Hamiltonian contains non-commuting terms, how the eigen-states are specified remains to be a basic problem in quantum mechanics.
In this paper, 2-species medium-body cold systems with spin-1 atoms (with particle numbers $N_A$ and $N_B$ being ranged from 8 to 60) are chosen as a model to study this basic problem, where the term arising from the inter-species interaction with the spins of the two associated different bosons coupled to $\lambda=1$ is non-commuting with other terms.
Making use of the fractional parentage coefficients, the associated Schrodinger equation has been solved exactly to obtain numerical solutions.
From them, we found that, although the total spin of the $A$-species $S_A$ and $S_B$ are no more conserved, but their averages $\overline{S_A}$ and $\overline{S_B}$ (i.e., their observables) are well defined.
Although they are not integers.
However, when the parameter for the strength of interaction varies, they vary in a step-by-step way (just like the behavior of a good quantum number).
Thus they can be used to specify the eigenstates.
Since the $\lambda=1$ channel has been mostly neglected in existing study, the present paper is a necessary complement to remedy this missing.
\end{abstract}


\maketitle

\section{Introduction}

In the field of ultracold physics Bose-Einstein condensates (BEC) are well known and have already been extensively studied. \cite{hos,esry97,fd1999,ib2008}
Both single-species systems \cite{stent98,ref5,ref6,ref6p,ref7,ref8,ref10,ref9} and multi-species systems \cite{pu98,ref17,man,sr2,ref14,ref17,hesre25} have been studied.
In the latter, due to the competition between the intra-species and inter-species interactions the spin textures become very rich. \cite{he3}
Thus these systems provide a platform for investigating interspecies interactions and complicated spin textures. \cite{acepjd,kegmprl,mlpra07,bjltp22}
Note that the spins of two different spin-$f$ bosons can be coupled to a combined spin $\lambda$ which is an even or odd integer.
The appearance of odd bi-species pairs is a distinctive feature in multi-species boson systems.
However, the effect of the odd-pairs has not yet been well studied previously.
We simply do not know how important these odd pairs would be.
Since the existence of the odd-pairs is extensive in multi-species boson systems.
This is a unacceptable missing and should be remedied.
As a necessary complement, this paper focuses on clarifying the effect of the odd pairs.

For this purpose, 2-species medium-body systems (with particle numbers ranged from 8 to 500) are chosen as the object of study.
Making use of the fractional parentage coefficient,\cite{bao05} a theoretical approach is proposed so that exact numerical solutions can be obtained.
Recall that there is a long-standing gap lying between the many-body theory and few-body theory.
As a complement, the medium-body theory could serve to fill this gap.
These systems are large enough to exhibit many-body behavior yet small enough to be treated rigorously.
The rigorous benchmarks provided by our exact solutions approve the accuracy of the physical results extracted in the following.

Note that, when the odd pairs have not been taken into account, the terms included in the Hamiltonian are commutative as shown in \cite{hesre25} and below.
Accordingly, there is a set of good quantum numbers to specify the eigen-states.
However, when the odd pairs are taken into account, an additional term emergences in the Hamiltonian (as shown below).
This term is not commuting with the other terms and leads to the vanish of the good quantum numbers.
How the eigen-states would be specified when the Hamiltonian contains non-commuting terms is a basic problem of quantum mechanics.
It turns out that our paper provides an exactly solvable model to dig out the answer of this basic problem.
This would be a topic interesting to all researchers in the field of quantum mechanics.

\section{Spin-dependent Hamiltonian of cold systems with two kinds of spin-1 atoms}

We consider a mixture of two kinds of spin-1 $X$-atoms ($X=A$ and $B$) with particle numbers $N_X$, bound in an optical trap.
The temperature $T$ is assumed to be sufficiently low (e.g., $T<10^{-10}$~K) \cite{cdprl21} so that all spatial degrees of freedom are frozen. \cite{hkpprl13,lzbj15}
In this case, the terms for the kinetic energy and potential energy (arising from the optical trap) are irrelevant and can be omitted.
Then the spin-dependent part of the Hamiltonian is \cite{mlpra07}
\begin{equation}
 \hat{H}
  =  \sum_{\lambda,i>j}
     g_{\lambda}^A
     P_{\lambda}^{A,i,j}
     \delta(\mathbf{r}_i-\mathbf{r}_j)
    +\sum_{\lambda,i>j}
     g_{\lambda}^B
     P_{\lambda}^{B,i,j}
     \delta(\mathbf{r}_i-\mathbf{r}_j)
    +\sum_{\lambda,i,i'}
     g_{\lambda}^{AB}
     P_{\lambda}^{AB,i,i'}
     \delta(\mathbf{r}_i-\mathbf{r}_{i'}),
\end{equation}
where $i$ and $j$ in the first (second) term denote two atoms of the $A$ ($B$)-species, while $i$ and $i'$ in the third term denote an $A$-atom and a $B$-atom, respectively.
$P_{\lambda}^{X,i,j}$ ($X=A$ or $B$) is the projector for extracting the component where the two spins of the $i$-th and $j$-th $X$-atoms are coupled to $\lambda=0$ or 2.
Similarly, $P_{\lambda}^{AB,i,i'}$ extract the spin-state where the $i$-th $A$-atom and the $i'$-th $B$-atom are coupled to $\lambda=0$, 1, or 2.
$g_{\lambda}^X$ ($g_{\lambda}^{AB}$) is the strength of the intra-$X$-species (inter-species) interaction related directly to the phase shift of the $\lambda$-channel.
Let the spatial wave functions be frozen at $\Phi_A$ and $\Phi_B$.
After integrating over all spatial degrees of freedom, we obtain the pure spin-dependent Hamiltonian
\begin{equation}
 \hat{H}_{\mathrm{spin}}
  =  \sum_{\lambda,i>j}
     \tilde{g}_{\lambda}^A
     P_{\lambda}^{A,i,j}
    +\sum_{\lambda,i>j}
     \tilde{g}_{\lambda}^B
     P_{\lambda}^{B,i,j}
    +\sum_{\lambda,i,i'}
     \tilde{g}_{\lambda}^{AB}
     P_{\lambda}^{AB,i,i'}
  \equiv
     \hat{H}_A
    +\hat{H}_B
    +\hat{H}_{AB},
 \label{h}
\end{equation}
where $\tilde{g}_{\lambda}^X=g_{\lambda}^X\int\Phi_X^4\mathrm{d}r$ and $\tilde{g}_{\lambda}^{AB}=g_{\lambda}^{AB}\int\Phi_A^2\Phi_B^2\mathrm{d}r$.

Note that there are a total of seven parameters $\tilde{g}_{\lambda}^Z$ ($Z=A$, $B$, and $AB$) contained in $\hat{H}_{\mathrm{spin}}$.
When the dimension of a parameter space is seven, the related analysis is very difficult.
Therefore, we are going to find out a much smaller parameter subspace without loss of physics.

Let $\hat{\mathbf{s}}_i$ and $\chi(i)$ be the spin operator and the spin-state of the $i$-th atom, respectively.
Note that, when the spin states of two atoms are coupled to $\lambda$, we have
\begin{equation}
 \hat{\mathbf{s}}_i
 \cdot
 \hat{\mathbf{s}}_j
 [\chi(i)\chi(j)]_{\lambda}
  =  q_{\lambda}
     [\chi(i)\chi(j)]_{\lambda},
 \label{ssq}
\end{equation}
with $q_0=-2$, $q_1=-1$, and $q_2=1$, where the $\lambda=1$ channel exists only if the two atoms are different as mentioned.
From Eq.(\ref{ssq}) and making use of the three identities, $\sum_{\lambda}P_{\lambda}^{Z,i,i'}=1$, we can establish a relation between the spin operators and the projectors.
It appears as $P_0^{Z,i,j}=\frac{1}{3}(1-\hat{\mathbf{s}}_i\cdot\hat{\mathbf{s}}_j)-2P_1^{Z,i,j}$ and $P_2^{Z,i,j}=\frac{1}{3}(2+\hat{\mathbf{s}}_i\cdot\hat{\mathbf{s}}_j)-P_1^{Z,i,j}$ (the last term in the above two formulae appears only if $Z=AB$).
Then, we find
\begin{eqnarray}
 \hat{H}_A
 &=& a\hat{S}_A^2
    +C_A, \label{ha} \\
 \hat{H}_B
 &=& b\hat{S}_B^2
    +C_B, \label{hb} \\
 \hat{H}_{AB}
  &=& 2c\hat{\mathbf{S}}_A
      \cdot
      \hat{\mathbf{S}}_B
     +C_{AB}
     +\sum_{i,i'}
      d
      P_1^{AB,i,i'}, \label{hab}
\end{eqnarray}
where $a=\frac{1}{6}(\tilde{g}_2^A-\tilde{g}_0^A)$, $b=\frac{1}{6}(\tilde{g}_2^B-\tilde{g}_0^B)$, $c=\frac{1}{6}(\tilde{g}_2^{AB}-\tilde{g}_0^{AB})$, $d=\frac{1}{3}(3\tilde{g}_1^{AB}-\tilde{g}_2^{AB}-2\tilde{g}_0^{AB})$, and $\hat{S}_X=\sum_{i\in X}\hat{\mathbf{s}}_i$ is the operator of the total spin of the $X$-species.
The three $C_Z$ are constants; they would only shift the spectra as a whole, but do not affect the details of the spin states, therefore, they can be dropped.
Thus, instead of seven, all physics remains in the 4-dimensional parameter space spanned by $a$, $b$, $c$, and $d$.

Furthermore, since the eigenstates do not depend on what energy unit is used.
Thus, when the norm of $c$ is used as the unit of energy, Eq.(\ref{h}) becomes
\begin{equation}
 \hat{H}_{\mathrm{spin}}/|c|
  =  (a'\pm 1)
     \hat{S}_A^2
    +(b'\pm 1)
     \hat{S}_B^2
    \mp
     \hat{S}^2
    +\sum_{i,i'}
     d'
     P_1^{AB,i,i'},
 \label{hc}
\end{equation}
where $S^2\equiv (S_A+S_B)^2$ is for the total spin of the whole system, $a'\equiv a/|c|$, $b'\equiv b/|c|$, and $d'\equiv d/|c|$.
For the signs $\pm$ or $\mp$ in the above equation, the upper one should be used if $c$ is negative.
Otherwise, the lower one should be used.
Thus, the dimension of the parameter space reduces further from seven to three.

The implications of these parameters are as follows.
Since the g.s. favors the texture with the lowest energy, from Eq.(\ref{ha}) we know that a negative $a$ (i.e., $\tilde{g}_2^A<\tilde{g}_0^A$) would cause the formation of 2-$A$-pairs (two $A$-atoms are coupled to 2) and these pairs would lie along the same direction to maximize $S_A$ so as to minimize $H_A$.
Whereas a positive $a$ (i.e., $\tilde{g}_0^A<\tilde{g}_2^A$) would promote the formation of singlet pairs (0-$A$-pair) to enable $H_A$ to be minimized.
Similarly, a negative $b$ tends to form the 2-$B$-pairs so as to maximize $S_B$, while a positive $b$ tends to form the 0-$B$-pairs so as to minimize $S_B$,

If the spins of all the atoms of a species are lying along the same direction, the species is called in the f-phase where all atoms form a fully polarized cluster.
Whereas if all the spins are 2-by-2 to form the 0-pairs, it is called in the p-phase.
If a species contains the 0-pairs together with a cluster (a polarized subsystem), it is called in the q-phase.
Thus, $a$ measures the ability to keep the $A$-species in f-phase (if it is negative) or in p-phase (if it is positive).
Similarly, $b$ measures the ability to keep the $B$-species in the f-phase or p-phase.
Besides, when the total spins of both species are nonzero, from Eq.(\ref{hab}) we know that a negative $c$ tends to maximize both $S_A$ and $S_B$, and pushes them lying along the same direction, while a positive $c$ tends also to maximize both $S_A$ and $S_B$ but pushes them lying along reverse directions.
Thus, $c$ affects not only the magnitudes but also the relative orientation of $S_A$ and $S_B$.
Thus, there is a competition between the intra- and inter-species interactions.
For example, when both $a$ and $b$ are positive, the intra-species interactions tend to keep both $S_A$ and $S_B$ as small as possible (in p-phase) while the interspecies interaction tends to make them as large as possible(in f-phase).
It turns out that the competition is in general crucial as shown below.

For carrying out the diagonalization of $\hat{H}_{\mathrm{spin}}/|c|$, it is necessary to introduce a complete set of basis states.
Note that, for a single $X$-species, let the associated normalized and symmetrized eigenstate be denoted as $\theta_{S_X}^{N_X}$ (where the total spin $S_X$ is a good quantum number ranging from $0$ to $N_X$, and $S_X$ should have the same even-odd parity as $N_X$).
It has been proved that the set $\{\theta_{S_X}^{N_X}\}$ is complete. \cite{ref10}
Therefore, the set $\{(\theta_{S_A}^{N_A}\theta_{S_B}^{N_B})_S\}\equiv\{\phi_{S_AS_BS}\}$ is complete for the binary system, where $S_A$ and $S_B$ are coupled to $S$ ranging from $0$ to $N_A+N_B$.
The detailed expression of $\theta_{S_X}^{N_X}$ is complicated.
However, due to the introduction of the fractional parentage coefficients, which have been derived in \cite{bao05} and given in Appendix I, all related 1-body and 2-body matrix elements of $\hat{H}_{\mathrm{spin}}/|c|$ can be easily obtained.
Thus, the detailed expression of $\theta_{S_X}^{N_X}$ is not necessary.

With this set and making use of the fractional parentage coefficients, the matrix elements of the Hamiltonian can be obtained.
They appear as
\begin{eqnarray}
 &&  \langle
     \phi_{S'_A S'_B S}|
     \hat{H}_\mathrm{spin}/|c||
     \phi_{S_A S_B S}
     \rangle \\
 &=& \delta_{S'_A S_A}
     \delta_{S'_B S_B}
     [ (a'\pm 1)
       S_A
       (S_A+1)
      +(b'\pm 1)
       S_B
       (S_B+1)
      \mp
       S(S+1)]  \nonumber \\ &&
      +d'
       N_A
       N_B
       \sum_{L J_A J_B I_A I_B}
       \mathfrak{U}(1 L S T_A^{\prime(J_A)}T_B^{\prime(J_B)})
       \mathfrak{U}(1 L S T_A^{(I_A)}T_B^{(I_B)})  \nonumber \\ && \ \ \ \ \ \ \ \ \ \ \ \ \ \ \ \ \ \ \ \ \ \ \ \ \times
       \delta_{T_A^{\prime(J_A)}T_A^{(I_A)}}
       \delta_{T_B^{\prime(J_B)}T_B^{(I_B)}}, \nonumber
 \label{hm}
\end{eqnarray}
where $L$ runs from $|1-S|$ to $1+S$, and all four indices $J_A$, $J_B$, $ I_A $, $I_B$, run from $1$ to $2$.
$T_A^{(1)}=S_A+1$, $T_A^{(2)}=S_A-1$, and $T_A^{\prime(1)}=S'_A+1$, $T_A^{\prime(2)}=S'_A-1$; these four definitions hold when the index $A$ is changed to $B$.
\begin{eqnarray}
   \mathfrak{U}(\lambda LST_A^{(I_A)}T_B^{(I_B)})
 &=& X_A^{(I_A)} X_B^{(I_B)} \\  &&\times
     \sqrt{(2\lambda +1)(2L+1)(2S_A+1)(2S_B+1)} \nonumber \\ &&\times
     U\left\{
     \begin{array}{ccc}
      1           & 1           &\lambda \\
      T_A^{(I_A)} & T_B^{(I_B)} & L \\
      S_A         & S_B         & S
     \end{array}
     \right\},  \nonumber  \label{u}
\end{eqnarray}
where the label $U$ with the nine indices is the 9-$j$ coefficient for the spins recoupling, and $X_A^{(1)}=a_{S_A}^{N_A}$, $X_A^{(2)}=b_{S_A}^{N_A}$, $X_B^{(1)}=a_{S_B}^{N_B}$, $X_B^{(2)}=b_{S_B}^{N_B}$, as referred to Eq.(\ref{fpc1}).

With this set of basis states $\{\phi_{S_AS_BS}\}$, after the diagonalization of the Hamiltonian, the exact eigenstates can be obtained.
Each can be expressed as a linear combination of the states in $\{\phi_{S_AS_BS}\}$.
In particular, the g.s. is denoted as
\begin{equation}
 \Psi_{\mathrm{gs}}
  =  \sum_{S_A S_B}
     \beta_{S_A S_B S}
     \phi_{S_AS_BS}.
 \label{gs}
\end{equation}
How it would be affected by the odd channel is analyzed in the following.

\section{A qualitative analysis of the effect of the odd channel on the ground state}

Before presenting the numerical data, we first study it in a qualitative way to better understand the inherent physics.
Recall that, making use of the fractional parentage coefficients given in the appendix, we can extract the $i$-th $A$-atom and the $j$-th $B$-atom from $\phi_{S_AS_BS}$ as
\begin{equation}
\phi_{S_AS_BS}=\sum_{\lambda LI_AI_B}\{[\chi^A(i)\chi
^B(j)]_{\lambda}(\theta_{T_A^{(I_A)}}^{N_A-1}\theta
_{T_B^{(I_B)}}^{N_B-1})_L\}_S\mathfrak{U}(\lambda
LST_A^{(I_A)}T_B^{(I_B)}),  \label{sasbs}
\end{equation}
where $\lambda=0$, $1$, and $2$, $L$ ranges from $|S-\lambda|$ to $S+\lambda$, and $I_A$ and $I_B$ both run from $1$ to $2$ as before.
Then, for the basis state $\phi_{S_AS_BS}$, the probability of the two specified particles forming a bi-species $\lambda$-pair is equal to
\begin{equation}
Q_{\lambda}^{S_AS_BS}\equiv\langle\phi_{S_AS_BS}|P_{\lambda
}^{AB,i,j}|\phi_{S_AS_BS}\rangle=\sum_{LI_AI_B}\mathfrak{U}
^2(\lambda LST_A^{(I_A)}T_B^{(I_B)}),  \label{qlam}
\end{equation}
[referring to Eq.(\ref{hm})].
Whereas for the g.s., the probability of forming a bi-species $\lambda$-pair is equal to
\begin{eqnarray}
Q_{\lambda}^{\mathrm{gs}} &\equiv &\langle\Psi_{\mathrm{gs}}|P_{\lambda
}^{AB,i,j}|\Psi_{\mathrm{gs}}\rangle  \nonumber \\
&=&\sum_{S_A'S_B'S_AS_B}\beta_{S'_AS_B'S}\beta_{S_AS_BS}\sum_{LJ_AJ_BI_AI_B} \\
&&\times\mathfrak{U}(\lambda LST_A^{\prime (J_A)}T_B^{\prime (J_B)})
\mathfrak{U}(\lambda LST_A^{(I_A)}T_B^{(I_B)})\delta_{T_A^{\prime(J_A)}T_A^{(I_A)}}\delta_{T_B^{\prime (J_B)}T_B^{(I_B)}}.
\nonumber  \label{qlamgs}
\end{eqnarray}

It turns out that, when the g.s. is in the p-p phase, i.e., $\Psi_{\mathrm{gs}}=\phi_{0,0,0}$, $Q_{\lambda}^{\mathrm{gs}}=Q_{\lambda}^{000}=\frac{2\lambda +1}{9}$.
Therefore, $Q_0^{000}=\frac{1}{9}$, $Q_1^{000}=\frac{3}{9}$, and $Q_2^{000}=\frac{5}{9}$ (This result arises from the isotropy).
On the other hand, when both $a'$ and $b'$ are negative and with $c>0$, the g.s. would be in the pure anti-parallel f-f phase, i.e., $\Psi_{\mathrm{gs}}=\phi_{N_AN_B|N_A-N_B|}\equiv\phi_{\mathrm{f-f}}$.
Let $N_0$ be the larger one of the pair $N_A$ and $N_B$, then we have $Q_0^{N_AN_B|N_A-N_B|}=\frac{2N_0+1}{3(2N_0-1)}\rightarrow\frac{1}{3}$, $Q_1^{N_AN_B|N_A-N_B|}=\frac{(N_0-1)(2N_0+1)}{2N_0(2N_0-1)}\rightarrow\frac{1}{2}$, and $Q_2^{N_AN_B|N_A-N_B|}=\frac{(N_0-1)(2N_0-3)}{6N_0(2N_0-1)}\rightarrow\frac{1}{6}$, where the limit is for the $N_0\rightarrow\infty$ case.
Whereas for parallel f-f phase, $\Psi_{\mathrm{gs}}=\phi_{N_AN_B(N_A+N_B)}$ (this happens when both $a'$ and $b'$ are negative and with $c<0$), then we have $Q_0^{N_AN_B(N_A+N_B)}=0$, $Q_1^{N_AN_B(N_A+N_B)}=0$, and $Q_2^{N_AN_B(N_A+N_B)}=1$.
Thus, we know that when each species itself is polarized (i.e., $S_X=N_X$), and $S=|N_A-N_B|$, the probability of forming the $\lambda=1$ bi-species pair is relatively larger, whereas if $S=N_A+N_B$, the probability of forming the $\lambda=1$ pairs becomes zero.

How the p-p$\rightleftarrows$f-f transition would be affected by the odd channel could be understood via the probability $Q_{\lambda}^{S_AS_BS}$.
When $c$ is negative, the parallel f-f phase would have $S=N_A+N_B$.
Note that, when $d'$ is negative, the spin state containing more $\lambda=1$ pairs will benefit more from the attractive $\lambda=1$ channel.
Since $Q_1^{000}>Q_1^{N_AN_B(N_A+N_B)}$, the p-p phase will benefit more than the f-f phase.
Thus, when $d'$ is negative, the transition p-p$\rightarrow$f-f would be hindered by $d'$, resulting in a larger domain of the p-p phase in the phase diagram.
Whereas when $d'$ is positive, the domain with the p-p phase would be contracted.
When $c$ is positive, the f-f phase would have $S=|N_A-N_B|$ .
Since $Q_1^{N_AN_B|N_A-N_B|}>Q_1^{000}$, an attractive odd channel would lower more energy of the anti-parallel f-f phase than that of the p-p phase.
Thus, when $d'$ is negative and $c>0$, the transition p-p$\rightarrow$f-f would be sped up by $d'$, resulting in a smaller domain of the p-p phase in the phase diagram.
Whereas when $d'$ is positive, the domain with the p-p phase would be enlarged.
This understanding could help us understand the following numerical results.

\section{Features of the ground state obtained via numerical calculation}

Once the diagonalization of the Hamiltonian has been carried out, we obtain the amplitudes $\beta_{S_AS_BS}$ of the g.s. (refer to Eq.(\ref{gs}).
We introduce two non-negative average values $\overline{S_X}$ ($X=A$ and $B$) that satisfy
\begin{equation}
\overline{S_X}(\overline{S_X}+1)\equiv\overline{S_X^2}=\langle\Psi_{\mathrm{gs}}|\hat{S}_X^2|\Psi_{\mathrm{gs}}\rangle
=\sum_{S_AS_B}\beta_{S_AS_BS}^2S_X(S_X+1).
\end{equation}
These two averages can be used to specify the character of the g.s. as shown below.

\subsection{The effect of $d'$ on the p-p phase}

We found that, when both $a'$ and $b'$ are positive and $a'b'$ is sufficiently large (refer to Ref.\cite{hesre25}), there is a broad domain in the $a'$-$b'$ parameter space where the g.s. has $S=0$ and $\overline{S_A}=\overline{S_B}$ is small (they are both exactly zero when $d'=0$).
We choose $a'=1.5$ and $b'=1.5$ as examples.
Then the amplitudes $\beta_{S_AS_A0}$ together with the associated $\overline{S_A}$ and $\overline{S_B}$ with different particle numbers under different $c'$ and $d'$ are shown in Table~\ref{tab1}.
The probabilities for any pair of different bosons forming a $\lambda$-pair in the g.s., $Q_{\lambda}^\mathrm{gs}$, together with the g.s. energy $E_\mathrm{gs}$, are also shown.

\begin{table}[!htb]
\caption{When $a'=1.5$ and $b'=1.5$, the g.s. is in p-p phase.
The features of this g.s. are demonstrated.
Only the five larger amplitudes $\beta_{S_A S_A 0}$ of the state have been given.
For each $d'$, three rows of data are given, the upper is for $N_A=12$, $N_B=8$, while both particle numbers are multiplied by 2 in the middle row, and multiplied by $5$ in the lower row.
$E_\mathrm{gs}$ is the g.s. energy, $Q_\lambda^\mathrm{gs}$ is the probability of forming a $\lambda$ bi-species pair [Eq.(\ref{qlamgs})].}
\label{tab1}
\begin{center}
\begin{tabular}{rrcccccccccc}
\hline\hline
\multicolumn{1}{c}{$c$} & \multicolumn{1}{c}{$d'$} & $E_\mathrm{gs}$
& $\beta_{000}$ & $\beta_{220}$ & $\beta_{440}$ & $\beta_{660}$ & $
\beta_{880}$ & $\overline{S_A}=\overline{S_B}$ & $Q_0^\mathrm{gs}$ & $Q_1^
\mathrm{gs}$ & $Q_2^\mathrm{gs}$ \\ \hline
$-1$ & $-2$ & $-85.3$ & $0.864$ & $-0.491$ & $0.110$ & $-0.011$ & $0.001$ & $
0.894$ & $0.014$ & $0.488$ & $0.498$ \\
&  & $(-361.2)$ & $(0.750)$ & $(-0.596)$ & $(0.273)$ & $(-0.081)$ & $(0.016)$
& $(1.543)$ & $(0.006)$ & $(0.496)$ & $(0.498)$ \\
&  & $(-2341)$ & $(0.608)$ & $(-0.593)$ & $(0.434)$ & $(-0.263)$ & $(0.134)$
& $(2.777)$ & $(0.002)$ & $(0.499)$ & $(0.499)$ \\
&  &  &  &  &  &  &  &  &  &  &  \\
$-1$ & $0$ & $0$ & $1$ & $0$ & $0$ & $0$ & $0$ & $0$ & $0.111$ & $0.333$ & $
0.556$ \\
&  & $(0)$ & $(1)$ & $(0)$ & $(0)$ & $(0)$ & $(0)$ & $(0)$ & $(0.111)$ & $
(0.333)$ & $(0.556)$ \\
&  & $(0)$ & $(1)$ & $(0)$ & $(0)$ & $(0)$ & $(0)$ & $(0)$ & $(0.111)$ & $
(0.333)$ & $(0.556)$ \\
&  &  &  &  &  &  &  &  &  &  &  \\
$-1$ & $2$ & $27.2$ & $0.704$ & $0.690$ & $0.166$ & $0.016$ & $0.001$ & $
1.416$ & $0.310$ & $0.053$ & $0.637$ \\
&  & $(55.6)$ & $(0.496)$ & $(0.752)$ & $(0.414)$ & $(0.128)$ & $(0.024)$ & $
(2.294)$ & $(0.324)$ & $(0.023)$ & $(0.652)$ \\
&  & $(140.2)$ & $(0.315)$ & $(0.605)$ & $(0.570)$ & $(0.393)$ & $(0.212)$ &
$(3.949)$ & $(0.330)$ & $(0.009)$ & $(0.661)$ \\
&  &  &  &  &  &  &  &  &  &  &  \\
$1$ & $-2$ & $-96.0$ & $0.716$ & $-0.611$ & $0.320$ & $-0.107$ & $0.020$ & $
1.747$ & $0.000$ & $0.525$ & $0.475$ \\
&  & $(-384.0)$ & $(0.618)$ & $(-0.598)$ & $(0.428)$ & $(-0.249)$ & $(0.116)$
& $(2.638)$ & $(0.000)$ & $(0.513)$ & $(0.488)$ \\
&  & $(-2400)$ & $(0.500)$ & $(-0.526)$ & $(0.460)$ & $(-0.368)$ & $(0.272)$
& $(4.424)$ & $(0.000)$ & $(0.505)$ & $(0.495)$ \\
&  &  &  &  &  &  &  &  &  &  &  \\
$1$ & $0$ & $0$ & $1$ & $0$ & $0$ & $0$ & $0$ & $0$ & $0.111$ & $0.333$ & $
0.556$ \\
&  & $(0)$ & $(1)$ & $(0)$ & $(0)$ & $(0)$ & $(0)$ & $(0)$ & $(0.111)$ & $
(0.333)$ & $(0.556)$ \\
&  & $(0)$ & $(1)$ & $(0)$ & $(0)$ & $(0)$ & $(0)$ & $(0)$ & $(0.111)$ & $
(0.333)$ & $(0.556)$ \\
&  &  &  &  &  &  &  &  &  &  &  \\
$1$ & $2$ & $8.2$ & $0.511$ & $0.760$ & $0.389$ & $0.099$ & $0.011$ & $2.176$
& $0.353$ & $0.007$ & $0.640$ \\
&  & $(16.2)$ & $(0.371)$ & $(0.672)$ & $(0.551)$ & $(0.303)$ & $(0.118)$ & $
(3.245)$ & $(0.343)$ & $(0.003)$ & $(0.654)$ \\
&  & $(40.2)$ & $(0.239)$ & $(0.490)$ & $(0.537)$ & $(0.471)$ & $(0.350)$ & $
(5.383)$ & $(0.337)$ & $(0.001)$ & $(0.661)$ \\ \hline\hline
\end{tabular}
\end{center}
\end{table}

\begin{table}[!htb]
\caption{$N_A=12$, $N_B=8$, $a'=-1.5$, $b'=-1.5$, $c<0$ (upper three rows) and $c>0$ (lower three rows), the features of the g.s. are demonstrated via the quantities listed in the table.
Only the three larger amplitudes $\beta_{S_A S_B S}$ of the g.s. have been given.}
\label{tab2}
\begin{center}
\begin{tabular}{rrcccccccccc}
\hline\hline
\multicolumn{1}{c}{$c$} & \multicolumn{1}{c}{$d'$} & $S$ & $E_
\mathrm{gs}$ & $\beta_{12,8,S}$ & $\beta_{10,8,S}$ & $\beta_{10,6,S}$ & $
\overline{S_A}$ & $\overline{S_B}$ & $Q_0^\mathrm{gs}$ & $Q_1^\mathrm{gs}$ &
$Q_2^\mathrm{gs}$ \\ \hline
$-1$ & $-2$ & $N_A+N_B$ & $-534.0$ & $1$ & $-$ & $-$ & $12$ & $8$ & $0$ & $0$
& $1$ \\
$-1$ & $0$ & $N_A+N_B$ & $-534.0$ & $1$ & $-$ & $-$ & $12$ & $8$ & $0$ & $0$
& $1$ \\
$-1$ & $2$ & $N_A+N_B$ & $-534.0$ & $1$ & $-$ & $-$ & $12$ & $8$ & $0$ & $0$
& $1$ \\
$1$ & $-2$ & $|N_A-N_B|$ & $-646.1$ & $0.999$ & $-0.007$ & $-0.048$ & $
11.996 $ & $7.996$ & $0.359$ & $0.503$ & $0.138$ \\
$1$ & $0$ & $|N_A-N_B|$ & $-550.0$ & $1$ & $0$ & $0$ & $12$ & $8$ & $0.362$
& $0.498$ & $0.139$ \\
$1$ & $2$ & $|N_A-N_B|$ & $-455.0$ & $0.998$ & $0.009$ & $0.059$ & $11.993$
& $7.994$ & $0.366$ & $0.492$ & $0.142$ \\ \hline\hline
\end{tabular}
\end{center}
\end{table}

\begin{table}[!htb]
\caption{The variation of $\overline{S_X}$ against $a'$ is demonstrated via $W_{Xi}$ and $(\widetilde{S_X})_i$ along the series of critical points, the case with $N_A=12$, $N_B=8$, $c<0$, and $b'=-1$ are assumed.}
\label{tab3}
\begin{center}
\begin{tabular}{rrccc}
\hline\hline
\multicolumn{1}{c}{$d'$} & \multicolumn{1}{c}{$i$} & $W_{Ai}$ & $
W_{Bi}$ & $[(\widetilde{S_A})_i,(\widetilde{S_B})_i]_S$ \\ \hline
$0$ & $0$ & $0$ & $0$ & $(12,8)_{20}$ \\
$0$ & $1$ & $0$ & $0$ & $(10,8)_{18}$ \\
$0$ & $2$ & $0$ & $0$ & $(8,8)_{16}$ \\
$0$ & $3$ & $0$ & $0$ & $(6,8)_{14}$ \\
$0$ & $4$ & $0$ & $0$ & $(4,8)_{12}$ \\
$-2$ & $0$ & $0$ & $0$ & $(12,8)_{20}$ \\
$-2$ & $1$ & $0.006$ & $-0.003$ & $(10.054,7.968)_{18}$ \\
$-2$ & $2$ & $0.019$ & $-0.009$ & $(8.126,7.933)_{16}$ \\
$-2$ & $3$ & $0.046$ & $-0.015$ & $(6.236,7.898)_{14}$ \\
$-2$ & $4$ & $0.041$ & $-0.005$ & $(4.448,7.876)_{12}$ \\
$2$ & $0$ & $0$ & $0$ & $(12,8)_{20}$ \\
$2$ & $1$ & $0.004$ & $-0.003$ & $(10.027,7.981)_{18}$ \\
$2$ & $2$ & $0.013$ & $-0.009$ & $(8.064,7.957)_{16}$ \\
$2$ & $3$ & $0.034$ & $-0.023$ & $(6.117,7.924)_{14}$ \\
$2$ & $4$ & $0.090$ & $-0.056$ & $(4.201,7.879)_{12}$ \\ \hline\hline
\end{tabular}
\end{center}
\end{table}

\begin{table}[!htb]
\caption{Similar to Table~\ref{tab3} but with $N_A=60$ and $N_B=40$.}
\label{tab4}
\begin{center}
\begin{tabular}{rrccc}
\hline\hline
\multicolumn{1}{c}{$d'$} & \multicolumn{1}{c}{$i$} & $W_{Ai}$ & $
W_{Bi}$ & $[(\widetilde{S_A})_i,(\widetilde{S_B})_i]_S$ \\ \hline
$0$ & $0$ & $0$ & $0$ & $(60,40)_{100}$ \\
$0$ & $1$ & $0$ & $0$ & $(58,40)_{98}$ \\
$0$ & $2$ & $0$ & $0$ & $(56,40)_{96}$ \\
$0$ & $3$ & $0$ & $0$ & $(54,40)_{94}$ \\
$0$ & $4$ & $0$ & $0$ & $(52,40)_{92}$ \\
$-2$ & $0$ & $0$ & $0$ & $(60,40)_{100}$ \\
$-2$ & $1$ & $0.001$ & $-0.001$ & $(58.043,39.961)_{98}$ \\
$-2$ & $2$ & $0.002$ & $-0.002$ & $(56.088,39.921)_{96}$ \\
$-2$ & $3$ & $0.003$ & $-0.003$ & $(54.135,39.880)_{94}$ \\
$-2$ & $4$ & $0.005$ & $-0.004$ & $(52.184,39.837)_{92}$ \\
$2$ & $0$ & $0$ & $0$ & $(60,40)_{100}$ \\
$2$ & $1$ & $0.001$ & $-0.001$ & $(58.023,39.978)_{98}$ \\
$2$ & $2$ & $0.001$ & $-0.001$ & $(56.048,39.955)_{96}$ \\
$2$ & $3$ & $0.002$ & $-0.002$ & $(54.074,39.931)_{94}$ \\
$2$ & $4$ & $0.003$ & $-0.003$ & $(52.101,39.906)_{92}$ \\ \hline\hline
\end{tabular}
\end{center}
\end{table}

\begin{table}[!htb]
\caption{Similar to Table~\ref{tab3} but with $b'=+1$, where, when $d'=0$, successive transitions $(12,8)_{20}\rightarrow(10,8)_{18}\rightarrow(0,0)_0$ are shown.
These transitions are slightly revised when $d'\neq 0$, where the p-p and f-f phases are not pure.}
\label{tab5}
\begin{center}
\begin{tabular}{rrccc}
\hline\hline
\multicolumn{1}{c}{$d'$} & \multicolumn{1}{c}{$i$} & $W_{Ai}$ & $
W_{Bi}$ & $[(\widetilde{S_A})_i,(\widetilde{S_B})_i]_S$ \\ \hline
$0$ & $0$ & $0$ & $0$ & $(12,8)_{20}$ \\
$0$ & $1$ & $0$ & $0$ & $(10,8)_{18}$ \\
$0$ & $2$ & $0$ & $0$ & $(0,0)_0$ \\
$-2$ & $0$ & $0$ & $0$ & $(12,8)_{20}$ \\
$-2$ & $1$ & $0.096$ & $-0.105$ & $(11.596,6.488)_{18}$ \\
$-2$ & $2$ & $0.305$ & $0.305$ & $(1.111,1.111)_0$ \\
$2$ & $0$ & $0$ & $0$ & $(12,8)_{20}$ \\
$2$ & $1$ & $0.045$ & $-0.037$ & $(10.193,7.846)_{18}$ \\
$2$ & $2$ & $0.061$ & $-0.053$ & $(8.623,7.492)_{16}$ \\
$2$ & $3$ & $0.240$ & $0.240$ & $(1.546,1.546)_0$ \\ \hline\hline
\end{tabular}
\end{center}
\end{table}

\begin{table}[!htb]
\caption{Similar to Table~\ref{tab5} but with the particle numbers $N_A=60$ and $N_B=40$.
When $d'=0$, multi-step successive transitions $(60-2i,40)_S\rightarrow(60-2i-2,40)_{S-2}$ emerge where $i$ is from $0$ to $9$ (those with $i>4$ are not shown).
Nonetheless, at the last critical point ($i=9$), the transition is a collective collapse $(42,40)_{82}\rightarrow(0,0)_0$.
When $d'\neq 0$, the multi-step successive transitions together with the big collapse remain, but the details are revised.}
\label{tab6}
\begin{center}
\begin{tabular}{rrccc}
\hline\hline
\multicolumn{1}{c}{$d'$} & \multicolumn{1}{c}{$i$} & $W_{Ai}$ & $
W_{Bi}$ & $[(\widetilde{S_A})_i,(\widetilde{S_B})_i]_S$ \\ \hline
$0$ & $0$ & $0$ & $0$ & $(60,40)_{100}$ \\
$0$ & $1$ & $0$ & $0$ & $(58,40)_{98}$ \\
$0$ & $2$ & $0$ & $0$ & $(56,40)_{96}$ \\
$0$ & $3$ & $0$ & $0$ & $(54,40)_{94}$ \\
$0$ & $4$ & $0$ & $0$ & $(52,40)_{92}$ \\
$-2$ & $0$ & $0$ & $0$ & $(60,40)_{100}$ \\
$-2$ & $1$ & $0.097$ & $-0.099$ & $(59.693,38.321)_{98}$ \\
$-2$ & $2$ & $0.120$ & $-0.122$ & $(59.090,36.943)_{96}$ \\
$-2$ & $3$ & $0.108$ & $-0.109$ & $(58.296,35.758)_{94}$ \\
$-2$ & $4$ & $0.089$ & $-0.089$ & $(57.386,34.689)_{92}$ \\
$2$ & $0$ & $0$ & $0$ & $(60,40)_{100}$ \\
$2$ & $1$ & $0.010$ & $-0.009$ & $(58.213,39.796)_{98}$ \\
$2$ & $2$ & $0.020$ & $-0.019$ & $(56.453,39.556)_{96}$ \\
$2$ & $3$ & $0.030$ & $-0.029$ & $(54.755,39.274)_{94}$ \\
$2$ & $4$ & $0.039$ & $-0.038$ & $(53.095,38.947)_{92}$ \\ \hline\hline
\end{tabular}
\end{center}
\end{table}

From this table, we see clearly that $d'$ causes fluctuation (the mixing of various $\phi_{S_AS_A0}$ components).
When $c<0$ and $d'<0$, the mixing is in a cyclic way (for any pair $\beta_{S_AS_A0}$ and $\beta_{S_A+2,S_A+2,0}$, they are always different in sign).
In this way, $Q_1^{\mathrm{gs}}$ becomes larger, resulting in a further decrease of $E_{\mathrm{gs}}$.
For example, when $c<0$ and $d'=-2$, the cyclic mixing causes an increase of $Q_1^{\mathrm{gs}}$ from $0.333$ to $0.488$ (if $N_A=12$ and $N_B=8$), and accordingly a decrease of $E_{\mathrm{gs}}$.
When $d'=-2$ and if there is no fluctuation, we have $\langle\phi_{0,0,0}|\hat{H}_{\mathrm{spin}}/|c||\phi_{0,0,0}\rangle=-64$.
However, for the real g.s. $\Psi_\mathrm{gs}$ with the fluctuation, we have Due to the fluctuation, we found $\langle\Psi_{gs}(d'=-2)|\hat{H}_{\mathrm{spin}}/|c||\Psi_{gs}(d'=-2)\rangle=-85.3$ as shown in the table.
Thus, the cyclic mixing causes an increase of $Q_1^{\mathrm{gs}}$ and therefore an additional deduction in $E_{\mathrm{gs}}$ by $21.3$).
Whereas when $c<0$ but $d'>0$, the mixing is in a coherent way (all $\beta_{S_AS_A0}$ have the same sign).
When $d'=2$, we found $\langle\phi_{0,0,0}|\hat{H}_{\mathrm{spin}}/|c||\phi_{0,0,0}\rangle=64$ and $\langle\Psi_{gs}(d'=2)|\hat{H}_{\mathrm{spin}}/|c||\Psi_{gs}(d'=2)\rangle=27.2$.
Thus the coherent mixing causes a decrease of $Q_1^{\mathrm{gs}}$ and therefore an additional decrease in $E_{gs}$ by 36.8.
In general, when $d'\neq 0$, the g.s. would adjust its structure (via different ways of fluctuating) to change $Q_1^{\mathrm{gs}}$ so as to increase the attraction or to reduce the repulsion from the odd channel.

The effect of particle numbers can be understood by comparing the data in the upper, middle, and lower rows for each pair of $c$ and $d'$.
A larger particle number will lead to a stronger fluctuation (a smaller $\beta_{000}$), and $\overline{S_A}=\overline{S_B}$ would have a larger deviation from zero.
However, the effect of particle number on $\{Q_{\lambda}^{\mathrm{gs}}\}$ is weak.
Besides, the way of coherence is not at all affected by the particle numbers.

\subsection{The effect of $d'$ on the f-f phase}

On the other hand, we found that, when both $a'$ and $b'$ are negative, the g.s. has $\overline{S_A}\simeq N_A$, and $\overline{S_B}\simeq N_B$ (they are exactly equal to $N_X$ when $c<0$ or $d'=0$) and $S=N_A+N_B$ (if $c<0$) or $S=|N_A-N_B|$ (if $c>0$).
In this case, each species is or is close to being fully polarized.
As an example, when $N_A=12$, $N_B=8$, $a'=-1.5$, $b'=-1.5$, the associated data are listed in Table~\ref{tab2}.
We found that, when $c<0$ and $S=N_A+N_B$, no $\lambda=1$ pairs would emerge in the parallel f-f phase (i.e., $Q_1^{\mathrm{gs}}=0$) because the formation of a $\lambda=1$ pair will spoil the conservation of $S=N_A+N_B$.
Therefore, $d'$ does not affect the parallel f-f state.
While for the anti-parallel f-f state with $S=|N_A-N_B|$, $Q_1^{\mathrm{gs}}\simeq\frac{1}{2}$ is large.
Accordingly, $E_{\mathrm{gs}}$ is greatly affected by $d'$ (say, when $c=1$, the introduction of the term $\sum_{i,i'}(-2)P_1^{AB,i,i'}$ in $\hat{H}_{\mathrm{spin}}/|c|$ would cause a decrease of $E_\mathrm{gs}$ from -550.0 to -646.1).
However, the associated fluctuation of amplitudes is found to be very slight, the shift of $S_X$ from $N_X$ is very slight.
and the change of $Q_1^{\mathrm{gs}}$ is also slight (from 0.498 to 0.503).
Thus, while the eigenstate is slightly changed, the eigenenergy is changed remarkably (because one more term is contained in the Hamiltonian).
Nonetheless, we see once again that different ways of coherence lead to a decrease or an increase of $Q_1^{\mathrm{gs}}$.

\begin{figure}[tbp]
\centering\resizebox{0.5\columnwidth}{!}{\includegraphics{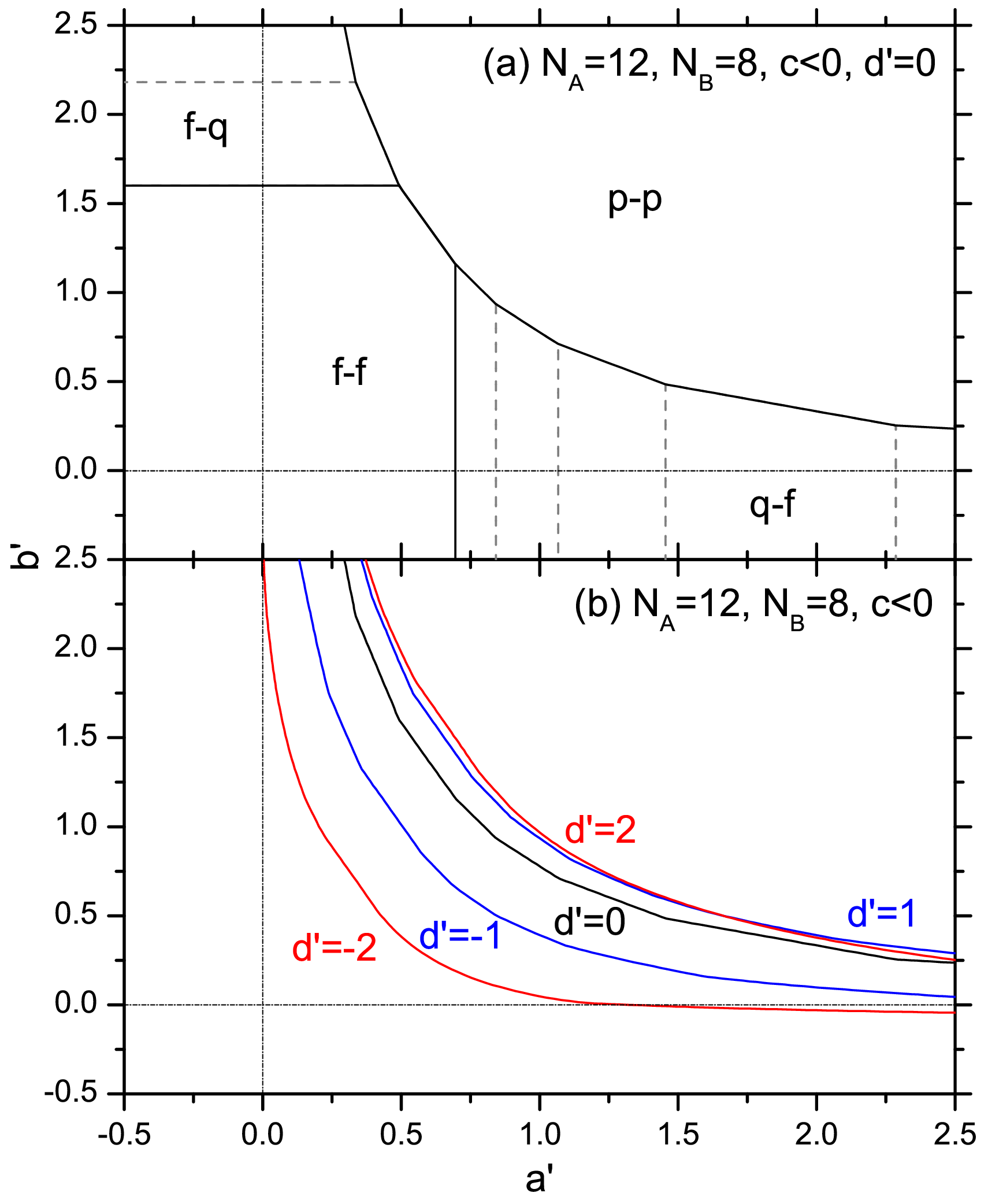}}
\caption{(color online) The phase diagram of the g.s. against $a'$ and $b'$, $N_A=12$, $N_B=8$, and $c<0$ are assumed.
(a) is for $d'=0$, where four domains for p-p, f-f, f-q, and q-f are marked.
Where f-q denotes that the $A$-species is in the f-phase while the $B$-species is in the q-phase, and so on.
In (b), $d'$ is given at five values (i.e., $-2$, $-1$, $0$, $1$, and $2$), and only the associated boundary between the p-p phase and other phases is plotted to demonstrate how the domains vary with $d'$.
The boundary is plotted in black if $d'=0$, in blue if $d'=\pm 1$, or in red if $d'=\pm 2$}.
\label{fig1}
\end{figure}

In Fig.\ref{fig1}a, the phase diagram of the g.s. is plotted against $a'$ and $b'$, where $c'<0$ and $d'=0$ are assumed.
There are two important neighboring domains with the p-p and f-f phases, respectively.
Based on the above discussion, we know that a negative $d'$ is favorable to the p-p phase and results in an enlargement of the p-p domain as shown in Fig.\ref{fig1}b.
Whereas a positive $d'$ would result in a contraction of the p-p domain.
In Fig.\ref{fig1}b, the p-p phase is so defined that $S=0$ and $\overline{S_A}=\overline{S_B}<2$.

\subsection{The variation of $\overline{S_X}$ against the parameters}

\begin{figure}[tbp]
\centering\resizebox{0.5\columnwidth}{!}{\includegraphics{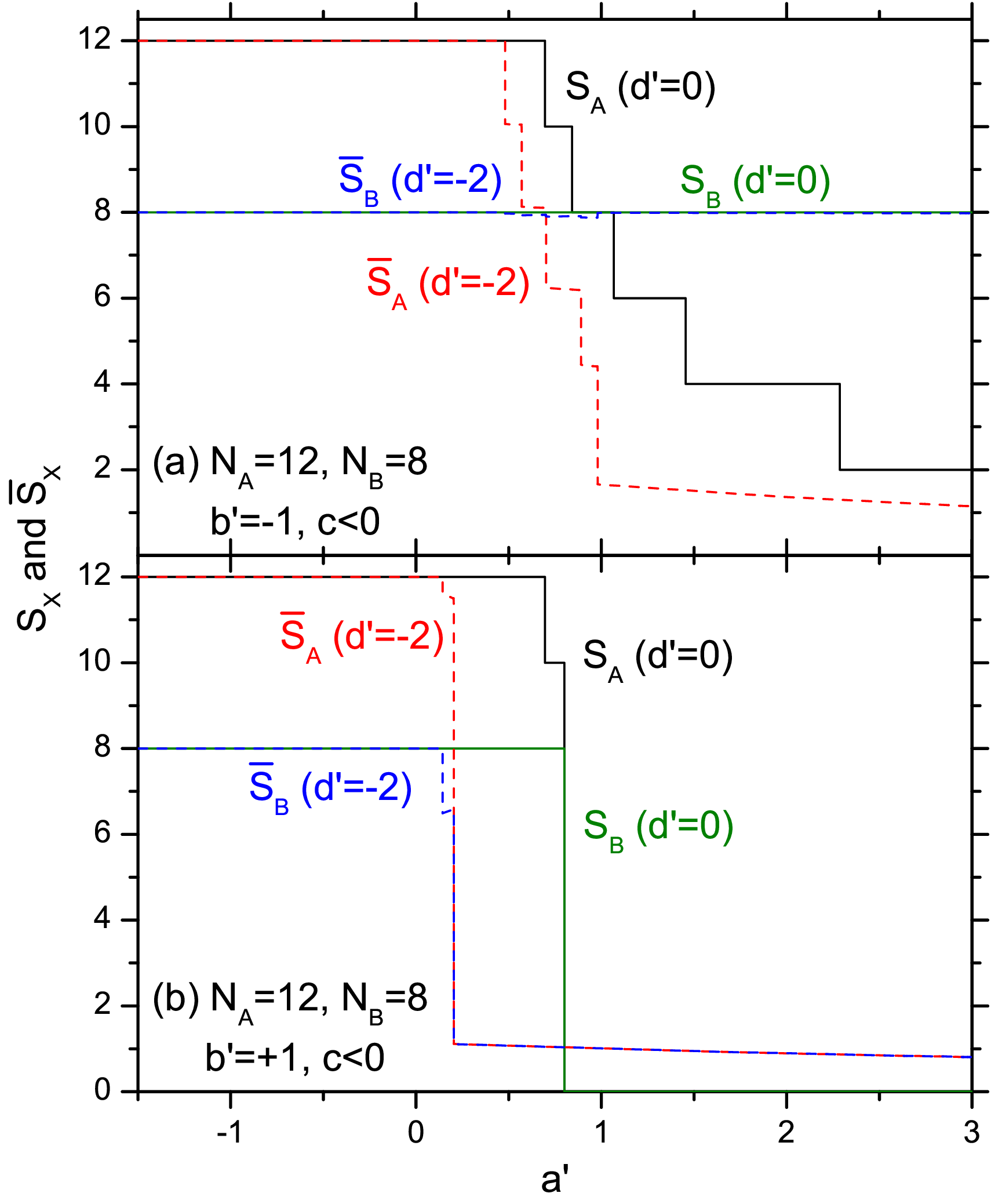}}
\caption{(color online) (a) When $N_A=12$, $N_B=8$, $c<0$ and $b'=-1$, the variation of $\overline{S_A}$ (in solid black line if $d'=0$ or in dashed red line if $d'=-2$) and $\overline{S_B}=N_B$ (in solid green line if $d'=0$ or in dashed blue line if $d'=-2$) against $a'$.
(b) Similar to (a) but with $b'=+1$.}
\label{fig2}
\end{figure}

\begin{figure}[tbp]
\centering\resizebox{0.5\columnwidth}{!}{\includegraphics{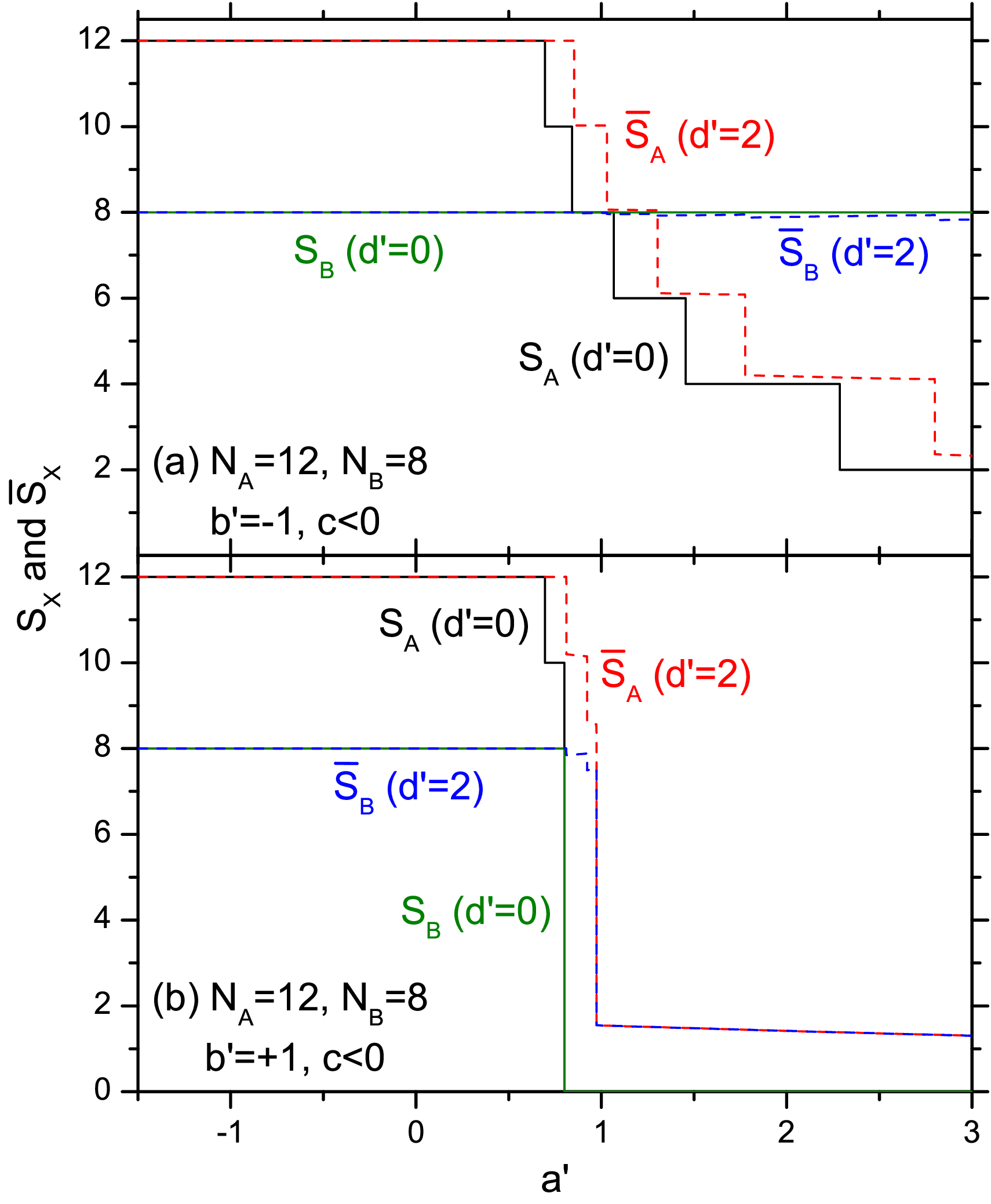}}
\caption{(color online) The same as Fig.\ref{fig2} but with $d'=0$ and $2$.}
\label{fig3}
\end{figure}

Recall that, when $d'\neq 0$, we do not have the good quantum numbers $S_X$ but $\overline{S_X}$.
Are these averages appropriate to specify an eigenstate? To clarify, the variations of $\overline{S_X}$ against the parameters are shown in Figs.\ref{fig2} and \ref{fig3}.

First, in Figs.\ref{fig2}a and \ref{fig3}a, $b'=-1$ is assumed, thus the intra-$B$-species interaction will attract the $B$-atoms lying along the same direction (i.e., in f-phase with $S_B=N_B$).
The solidity of the f-phase of the $B$-species arises from the fact that, if the $A$-atoms are also in f-phase, then the inter-species interaction could only affect the relative orientation of $S_A$ and $S_B$ but not their values.
If some $A$-atoms are in 0-pairs, the inter-species interaction arising from these 0-pairs is mutually canceled because of the isotropism of the 0-pairs.
This leads to the solidity.
Similarly, a negative $a'$ would also lead to the f-phase.
Whereas a positive $a'$ would lead to the formation of the 0-pairs, thus the increase of $a'$ would lead to the appearance of more 0-$A$-pairs and therefore a decrease of $\overline{S_A}$ as shown in the figures.
When $d'=0$, the decrease of $S_A$ is in a step-by-step way, in each step, the decrease of $S_A$ is exactly $2$.
It is notable that, when $d'\neq 0$, the decrease of $\overline{S_A}$ is also in a step-by-step way.
There are some critical points (labeled by the index $i$) where a fall of $\overline{S_A}$ occurs.
For example, for the case $d'=-2$ (Fig.\ref{fig2}a), when $a'$ increases and arrives at the first critical point ($i=1$) located at $a'=0.480$, a transition occurs and $\overline{S_A}$ falls suddenly from $12$ to $10.054$.
Then, when $a'$ increases in an interval between two critical points, $\overline{S_A}$ decreases extremely slowly until it arrives at the next critical point, where the second fall occurs, and so on.
In general, at the $i$-th critical point $a_i'$ where a fall of $\overline{S_A}$ occurs, let the value of $\overline{S_A}$ at the top (bottom) of the fall be denoted as $(\overline{S_A})_{i,\uparrow}$ ($(\overline{S_A})_{i,\downarrow}$).
Then the variation of $\overline{S_A}$ can be specified by the critical points together with the height of the fall $h_{Ai}\equiv[(\overline{S_A} )_{i,\uparrow}-(\overline{S_A})_{i,\downarrow}]$ and the width of the $i$-th interval $W_{Ai}\equiv[(\overline{S_A})_{i,\downarrow}-(\overline{S_A})_{i+1,\uparrow}]$.
It was found that, when $a'$ increases in a reasonable scope (not too larger than $b'$), $h_{Ai}\simeq 2\gg W_{Ai}$.
Since $W_{Ai}$ is so small, $\overline{S_A}$ seems to vary disruptively and jump from a constant (a real number) $(\overline{S_A})_{i,\uparrow}$ to the next real number $(\overline{S_A} )_{i,\downarrow}\simeq(\overline{S_A})_{i+1,\uparrow}$.

The cases with a positive $d'$ are similar.
However, comparing Fig.\ref{fig2}a with Fig.\ref{fig3}a, the curve for $\overline{S_A}$ with $d'=-2$ shifts to the left from the one with $d'=0$, implying that the attractive odd channel benefits the formation of the 0-pairs as mentioned above, while the curve with $d'=2$ shifts to the right.
Furthermore, for the $B$-species with $b'=-1$, the f-phase is so solid that the variation of $\overline{S_B}$ against $a'$ is so small that it cannot be seen in the figures.
The details are shown in Table~\ref{tab3}.

From this table and Table~\ref{tab4} (where the particle numbers are enlarged by a factor of $5$), we see that, when $N_X$ is larger, the widths are really very narrow and the variation of $\overline{S_A}$ against $a'$ is more likely in a step-by-step way.
This picture would be better established when $i$ is smaller.
Thus, for a many-body system with reasonable strengths (say, $a'$ and $b'$ are not remarkably different in magnitudes), the above picture would hold nicely.

The case with $b'=+1$, as shown in Figs.\ref{fig2}b and \ref{fig3} b, and in Tables~\ref{tab5} and \ref{tab6}, is very different.
When $b'>0$, $a'<0$, and $c=0$, the $B$-species alone would prefer the p-phase while the $A$-species would prefer the f-phase.
However, when $c<0$, we found that, while the $A$-species remains in the f-phase, the $B$-species is also in the f-phase even when $b'>0$.
This is because all the spins of the $A$-atoms are parallel, the attraction acting upon each $B$-atom from them (via the negative $c$) can be mutually enhanced and henceforth is sufficient to push every $B$-atom lying along the same direction as them.
Thus, the attraction would lead to the f-f phase as shown by the left part of the black and green curves in Fig.\ref{fig2}b.
Even in the case with both $b'>0$ and $a'>0$, if $|c|$ is large enough so that the product $a'b'$ is small enough, the g.s. would still keep itself in the f-f phase.
In this way, the attraction arising from the inter-interactions can be maximized (refer to Ref.\cite{hesre25}).
Accordingly, there are two crucial points.
We see in Fig.\ref{fig2}b that once $a'$ becomes more repulsive and exceeds the first crucial point, a collapse of the f-f phase together with the transition f-f$\rightarrow$q-f phase [i.e., $(N_A,N_B)_S\rightarrow (N_A-2,N_B)_{S-2}$] occurs.
When $a'$ crosses over the second crucial point, the transition q-f$\rightarrow$p-p [$(N_A-2,N_B)_{S-2}\rightarrow (0,0)_0$] occurs successively.
If $a'>0$ and $b'>0$ are both sufficiently large, we found only one crucial point associated with the collective transition f-f$\rightarrow$p-p.
Thus, a more repulsive $b'$ helps the $B$-atoms to form the 0-$B$-pairs leading to the vanish of the polarized $B$-cluster and therefore favors the appearance of the 0-$A$-pairs.
Thus, more 0-$B$-pairs would lead to more 0-$A$-pairs, and vice versa This mechanism leads to the collective transition Ref.\cite{hesre25}.
In fact, we found in the phase-diagram that the f-f and p-p phases are neighboring.
The details with $N_A=12$ and $N_B=8$ are given in Table~\ref{tab5}, while the details with $N_A=60$ and $N_B=40$ are given in Table~\ref{tab6}.

Comparing Table~\ref{tab5} with Table~\ref{tab3}, we found that a positive $b'$ would lead to a broader width.
However, we know from Table~\ref{tab6} that there is still a region in the parameter space where the intervals remain narrow when the particle numbers are larger (say, for the case with $d'=2$, $b'=+1$, and $i=4$ given in Table~\ref{tab6}).

\section{Final remarks}

We have found that the g.s. energy $E_{\mathrm{gs}}$ is directly affected by the odd channel.
Accordingly, the phase diagrams are modified by $d'$ as shown in Fig.\ref{fig1}.
Besides, the odd channel causes an mixing of various components $\phi_{S_AS_BS}$ in every eigenstate.
Accordingly, $S_X$ is no longer conserved (while the total spin $S$ is).
Two kinds of coherence are found in the mixing: coherent mixing and cyclic mixing.
For the g.s., the ways of mixing aim at increasing $Q_1^{\mathrm{gs}}$ to strengthen the attraction from a negative $d'$, or at decreasing $Q_1^{\mathrm{gs}}$ to reduce the repulsion from a positive $d'$.
Since the odd channel exist in all multi-species boson systems and the effect of this channel is explicit, the study of this channel is a necessary complement.

Due to the mixing the previous good quantum numbers $S_A$ and $S_B$ are replaced by $\overline{S_A}$ and $\overline{S_B}$.
Just as the former two, the latter two vary with the parameters also in a step-by-step way.
They jump from a narrow interval suddenly to the next well-separated narrow interval.
These narrow intervals are so narrow, thus each interval can be considered as a real number, and can be called a nearly good quantum number to specify the g.s..

The above study provides an example for clarifying what will happen when a Hamiltonian (containing commuting terms) is interfered by an additional non-commuting term.
Obviously, this is a basic problem in Quantum mechanics.
Since we have obtained exact solutions, the physics extracted from them is accurate.
However, for various many-body systems and for various excited states, how general the above finding would be remains to be clarified.
What is particularly interesting is whether the width of the narrow interval tends to zero when the number of particles approaches infinity?

Finally, it is worthy to mention that the medium-body systems are scarcely studied due to the difficulty in theoretical treatment.
Nonetheless, the introduction of the fractional parentage coefficients facilitate greatly the calculation of the matrix elements of the whole Hamiltonian.
Nonetheless, the dimension of the matrix is crucial.
When the spatial degrees of freedom have been frozen, the dimension is determined by the number of basis-states contained in the set $\{\phi_{S_AS_BS}\}$ with a given $S$.
For a medium-body 2-species system as large as having  $N_A=300$ and $N_B=200$ , the largest number of basis-states is 9334, it happens when $S=166$.
Thus we know that for general medium-body ultracold systems with spin-1 bosons, our approach is applicable

The medium-body theory can serve as a bridge to relate many-body physics and few-body physics.
In particular, the rigorous solutions of medium-body systems allow for a stringent test of many-body approximations and clarify which few-body phenomena persist in the thermodynamic limit.
Thus the study of these systems is worthy.

\section*{Competing Interests}

The authors declare no competing interests.

\section*{Data Availability}

All data generated or analyzed during this study are included in this published article.

\section*{Funding Declaration}

No funds have been received for this manuscript.

\section*{Author Contributions}

Yanzhang He is responsible to the theoretical derivation, numerical calculation, and the whole paper.
Yimin Liu is responsible to the theoretical derivation.
Chengguang Bao provides the idea and writes the paper.
All authors reviewed the manuscript.

\section*{Appendix: Fractional parentage coefficients of spin-1 systems}

We consider a spin-1 system containing $N$ particles of the same species governed by the Hamiltonian $\sum_{\lambda,i>j}\tilde{g}_\lambda P_\lambda^{i,j}$.
The total spin $S$ and its $Z$-component $M$ are conserved.
Thus, the normalized and symmetrized eigenstates can be specified by $S$ and $M$ as $\theta_{SM}^N$, where $N-S$ must be even; otherwise, $\theta_{SM}^N$ is zero.
It has been proved that the multiplicity of $\theta_{SM}^N$ is one. \cite{katr}
Thus, $\theta_{SM}^N$ is unique, and the set $\{\theta_{SM}^N\}$ is complete.
One can extract a particle, say, the $i$-th particle, from this state.
After the extraction, the total spin of the rest part must be either $S+1$ or $S-1$.
The other choice is not possible due to the even-odd parity.
Thus, we have
\begin{equation}
\theta_{SM}^N=a_S^N [\chi(i)\theta_{S+1}^{N-1}]_{SM} +b_S^N
[\chi(i)\theta_{S-1}^{N-1}]_{SM},  \label{fpc1}
\end{equation}
where $\chi(i)$ is the spin state of the $i$-th particle.

There has already been a study on the coefficients in this equation.
It has been proven that \cite{bao05}
\begin{equation}
a_S^N=[\frac{(N-S)(S+1)}{N(2S+1)}]^{1/2}, \ \ \ b_S^N=[\frac{(N+S+1)S}{
N(2S+1)}]^{1/2}.
\end{equation}
They are called the 1-body fractional parentage coefficients.

For convenience, Eq.(\ref{fpc1}) can be rewritten as
\begin{equation}
\theta_{SM}^N=\sum_{S'} a_{S'S}^N
[\chi(i)\theta_{S'}^{N-1}]_{SM},  \label{fpc1p}
\end{equation}
where
\begin{equation}
a_{S'S}^N=[\frac{(N-S)(S+1)}{N(2S+1)}]^{1/2}\delta_{S',S+1} +[\frac{(N+S+1)S}{N(2S+1)}]^{1/2}\delta_{S',S-1}.
\end{equation}
When one more particle, say, the $j$-th particle, is further extracted from Eq.(\ref{fpc1}), we have
\begin{equation}
\theta_{SM}^N=\sum_{\lambda,S'} h_{\lambda S'S}^N
\{[\chi(i)\chi(j)]_\lambda\theta_{S'}^{N-2}\}_{SM},  \label{fpc2}
\end{equation}
where $\chi(i)$ and $\chi(j)$ are coupled to $\lambda$.
They both have been extracted, and
\begin{eqnarray}
h_{0,S,S}^N &=& [\frac{(N-S)(N+S+1)} {3N(N-1)}]^{1/2}, \\
h_{2SS}^N &=& [\frac{S(2S+2)(N-S)(N+S+1)} {3(2S-1)(2S+3)N(N-1)}]^{1/2}, \\
h_{2,S+2,S}^N &=& [\frac{(S+1)(S+2)(N-S)(N-S-2)} {(2S+1)(2S+3)N(N-1)}]^{1/2},
\\
h_{2,S-2,S}^N &=& [\frac{S(S-1)(N+S+1)(N+S-1)} {(2S-1)(2S+1)N(N-1)}]^{1/2},
\end{eqnarray}
and $h_{\lambda S'S}^N=0$, if $S'$ does not belong to the above cases.

The coefficients in Eq.(\ref{fpc2}) are called 2-body fractional parentage coefficients.

With the above formulae and coefficients, the matrix elements of any 1-body and 2-body operators against the eigenstates can be conveniently and analytically derived.
This is a great advantage because all relevant physical quantities of the system can be obtained without knowing the details of $\theta_{SM}^N$.

\end{document}